\documentclass[page-classic]{eplv} 

\title{Assessment of the beta-delayed proton decay rate of $^{11}$Be}
\shorttitle{Beta-delayed proton decay of $^{11}$Be} 

\author{A. Volya\inst{1,2}}
\shortauthor{A. Volya \etal}

\institute{                    
  \inst{1} Department of Physics, Florida State University, Tallahassee, FL 32306,
USA\\
  \inst{2} Cyclotron Institute, Texas A\&M University, College Station, Texas 77843, USA
}

\abstract{
The $^{11}$Be neutron halo nucleus appears to decay into $^{10}$Be with a rate that exceeds expectations. Neutron disappearance into dark matter, beta decay of a halo neutron, or beta-delayed proton decay have been offered as explanations. In this work we study the latter option; we carry out shell model calculations and sequential decay analysis examining the beta-delayed proton decay going through a resonance in $^{11}$B. The narrow energy window, lack of states with sufficient spectroscopic strength, overwhelming
alpha decay branch, all make reconciling the observed rate with beta-delayed proton decay difficult. 
}

\begin{document}

\maketitle

\section{Introduction}
The unexplained difference in the observed neutron lifetime  between two types of experiments \cite{wietfeldt:2011} has fueled multiple speculations including suggestions that this difference is due to a decay into some dark matter particle(s). 
This idea is examined in Ref. \cite{pfutzner:2018} where the possibility to assess this decay branch using other nuclear beta decays is addressed. The nucleus $^{11}$Be is identified as one of the best candidates; $^{11}$Be with four protons 
and seven neutrons is known as a neutron halo nucleus. Its structure is rather interesting: the six neutrons occupy  the lowest 
$0s_{1/2}$ and $0p_{3/2}$ shells of the mean-field potential and the remaining halo neutron is on the $1s_{1/2}$ shell which is just 320 keV below $0p_{1/2}.$ This modification of the shell structure appears to be a result of strong continuum coupling of the $s$-orbit 
\cite{volya:2014ad:proc,volya:2012aa:proc,hoffman:2014} which is pushed down by the virtual excitations. 
Therefore, in terms of the harmonic oscillator quantum numbers, the two-quanta $1s_{1/2}$ state appears below the one-quantum $0p_{1/2}$ state. The neutron separation energy in $^{11}$Be is only 0.5016 MeV making the $s$-wave orbit geometrically large. Thus, even without introducing any many-body complexity, one could envision a halo neutron undergoing a decay process, similar to that of a free neutron decay \cite{baye:2011}. This decay could also have a dark matter branch of interest. 

The neutron beta decay
\begin{equation}
n\rightarrow p + e^{-} + \bar{\nu}_e
\end{equation}
releases 0.782 MeV of kinetic energy (decay $Q$-value) that is shared between the products. The exact number is uncertain due to a small mass of the neutrino but this uncertainty is too small to be relevant here. The neutron half-life $t_n$ is likely
to be between 610 and 613 seconds were uncertainty possibly comes from a dark matter decay branch, Ref. \cite{wietfeldt:2011,pfutzner:2018,ejiri:2019}. 
This uncertainty can be used to make estimates of the dark matter branch where it is hypothesized that neutron can also
decay to some unknown fragments $X.$  The corresponding half-life of dark decay could be
$t_{n\rightarrow X}=6.8\times 10^4$s, see Refs. \cite{pfutzner:2018,fornal:2018}.
The $Q$-value in this decay can range from 0 up to 1.572 MeV. The upper limit is restricted by the fact that a neutron in $^9$Be does not decay so the $Q$-value is less then the energy required for breakup $^9{\rm Be}\rightarrow n+\alpha+\alpha,$ see  Ref. \cite{pfutzner:2018}.

The neutron decay is a classic example of the allowed beta decay process. The beta-decay half-life $t$
is typically expressed using $ft$ value,  see for example textbook \cite{zelevinsky:2017}, 
\begin{equation}
f t=\frac{\cal T}{B(F)+\lambda_A^2 B (GT)} .
\label{eq:ft}
\end{equation}
The expression includes the energy dependent phase space function $f(\epsilon)$ evaluated at the $Q$-value, constant ${\cal T}=6145$ s, and reduced matrix elements for Fermi $B(F)$ and Gamow-Teller $B(GT)$ operators; the $B(GT)$ comes with a coupling constant $\lambda_A=1.27.$ The expression in eq. (\ref{eq:ft}) reproduces the neutron half-life given that $B(F)=1,$ $B(GT)=3,$
and $f(Q)=1.715$ The beta decay width as a function of decay energy,
\begin{equation}
\gamma_\beta(\epsilon) = \frac{\ln 2}{t}= \frac{f(\epsilon)\,\hbar\, \ln 2 }{\cal T}\left (B(F)+\lambda_A^2 B (GT)\right )
\label{eq:gb}
\end{equation}
follows the Fermi's Golden rule applicable to decay processes in general,  where energy dependence is determined by the corresponding phase space volume with the remaining multiplier being comprised of coupling constants and transition matrix elements. The phase space volume grows very fast as a function of energy, typically as a power law, and for this reason decay rates are very sensitive to the $Q$-values. 

The beta decay 
\begin{equation}
^{11}{\rm Be} \rightarrow ^{11}{\rm B} + e^{-} + \bar{\nu}_e
\end{equation}
releases 11.5092 MeV of kinetic energy because a proton is strongly bound in $^{11}$B, thus leading to a much shorter half-life
$t_{{\rm Be}}=13.76$ seconds. At the same time the $Q$-value for the dark matter decay is reduced by the neutron separation energy.
The phase space arguments can be used to give a lower limit for the half-life for dark matter decay $t_{{\rm Be}\rightarrow X}> 10^5$ s,
see discussion of the estimates in 
Ref. \cite{pfutzner:2018}.


The beta decay rate of  $^{11}$Be is comprised of many decay branches populating excited states in $^{11}$B, the ground state branching ratio is only about 55\%. Most of the remaining decay proceeds into excited states of 
$^{11}$B, some of those states further decay by particle emissions. 
The beta-delayed alpha emission $Q_{\beta\alpha}=2.845$ MeV is known to have a
3.3$\%$ branch, see also Ref. \cite{refsgaard:2019}.
The beta-delayed neutron emission  $Q_{\beta n}=0.0551$ MeV has not been observed. Finally, 
the beta-delayed proton emission $Q_{\beta p}=0.2807$ MeV is of a particular interest because it is indistinct from the 
halo neutron beta decay that, as discussed earlier, can also include a dark matter branch. 

Several experiments have been carried out \cite{borge:2013,riisager:2014,ayyad:2019} determining the branching ratio for the beta-delayed proton emission
at $1.3\times 10^{-5}$ which corresponds to half-life for this decay branch $t_{{\rm Be}\rightarrow \beta p}\approx1\times 10^6$ s. 
%
The phase-space-based estimate for the halo-neutron beta decay carried out in Ref. \cite{ayyad:2019} gives  
$t_{{\rm Be}\rightarrow \beta p}= 2.2\times10^{10}$ s. A more detailed calculation in Ref \cite{baye:2011} gives $t_{{\rm Be}\rightarrow \beta p} \sim 10^{8}$ s.
Both estimates suggest half-lives that are several orders of magnitude longer than the observed value. 
The significantly enhanced rate of $^{11}$Be decay into $^{10}$Be fueled the dark matter related speculations; in particular because
the observed half-life agrees well with the estimates discussed above and  in Ref. \cite{pfutzner:2018}.

As a non-exotic explanation,
the authors of Ref. \cite{ayyad:2019} suggest a resonance in $^{11}$B at 196 keV from the proton decay threshold; and using the experimental proton energy distribution they estimate the width of the proton resonance to be $\Gamma=12(5)$ keV, and 
$\log_{10}(ft)=4.8(4)$ is suggested for the beta decay into this resonance. 
%
While a beta-delayed proton emission going through a resonance in $^{11}$B is a possible explanation, there are some reasons for 
skepticism. 

\begin{figure}[h]
\begin{center}
\includegraphics[width=0.8\linewidth]{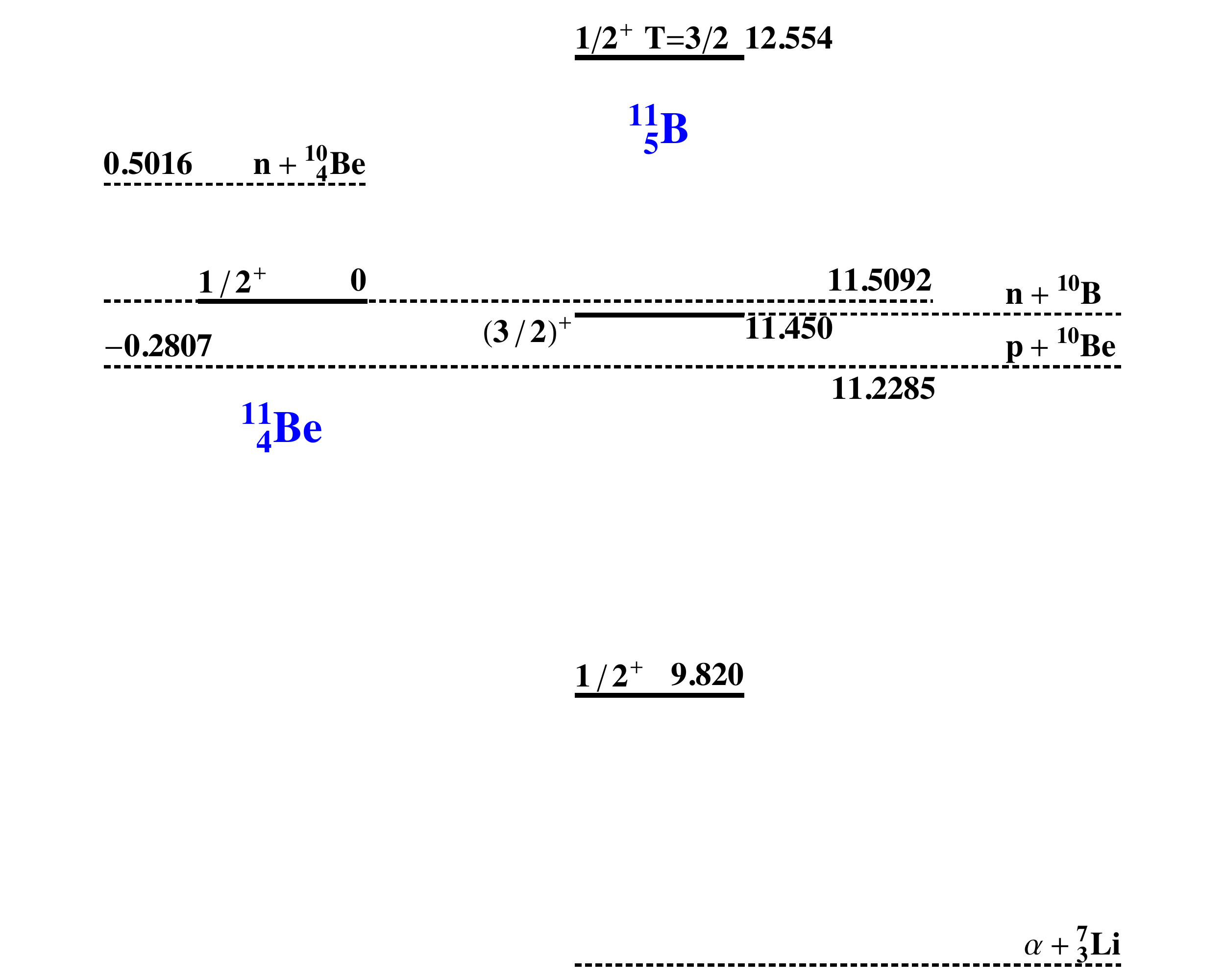}
\end{center}
\caption{\label{fig:spec} Level structure diagram of $^{11}$B between 8.6 and 12.6 MeV of excitation. 
}
\end{figure}

In this work we analyze the beta-delayed proton emission using detailed microscopic shell model calculations and full analysis of the sequential decay process. The proton separation energy in $^{11}$B is at $11.2285$ MeV of excitation and the $^{11}$Be threshold is 280.7 keV
above it at 11.5092 MeV. So a proton resonance within this energy range could allow for a cascading beta-delayed decay. 
The level structure of $^{11}$B in the energy region of interest is shown in Fig. \ref{fig:spec}.
The experimentally known levels with spin-parity $1/2^+$ and $3/2^+$ that can be populated by allowed beta decay from 
the $1/2^+$ ground state of $^{11}$Be are shown. 
The dashed line at -0.2807 MeV shows the beta-delayed proton emission threshold. The nucleus of $^{10}$Be has a $J^{\pi}=0^{+}$ ground state and since the first excited state is above 3 MeV in excitation only beta-delayed proton decay to the ground state of $^{10}$Be is possible. The threshold for alpha decay is 2.8 MeV lower, and this channel is open for all states in the region.
%

\section{Shell model analysis}
Assessing the possibility for the beta-delayed decay proceeding through a resonance in $^{11}$B we 
carried out shell model calculations using $p-sd$ shell model space 
with {\it psdu} interaction from Ref. \cite{utsuno:2011} where shell gap between $p$ and $sd$ shells was modified reproduce relative positions of negative and positive parity states. The beta decay rate observed in Ref. \cite{ayyad:2019} indicates an allowed transition
that has to be relatively strong; the proton decay that follows has to be strong as well in order to compete with the alpha channel. 
Thus, the resonant state(s) of interest in the narrow energy window above 11.2 MeV of excitation has to have a relatively simple 
structure and our shell model approach should be able to describe it. 
As a reference, we also carried out calculations using {\it fsu} interaction from Ref. \cite{lubna:2019:art} which only includes one particle-hole excitation but describes well structure of nuclei across several shells. 

In Tab. \ref{tab:1} we list all states of negative parity in $^{11}$B up to experimentally known fifth $5/2^{-}$  
state (labeled as $5/2^{-}_5$), and all positive parity state up to $3/2^+_5.$
While the predicted shell model energies and experimentally observed energies are slightly different, they are well within the typical 
shell model errors,  and the overall low-lying sequence of levels is reproduced. 
There is a complete one-to-one agreement up to $7/2^+_2$ at 10.597 MeV. 
We will not discuss the negative parity states since they would not be populated in the allowed beta decay. 
In the region of interest, above $9/2^+_1$ at 11.265 MeV and  below $7/2^+_3$ at 12 MeV,  the shell model predicts two
positive parity states $9/2^+_2$ and $3/2^+_3.$ 
In the table we correlate a state  at 11.45 MeV of unknown spin-parity with $3/2^+_3.$  
The table also includes shell model predictions up to the following two $3/2^+$ states and one $1/2^+$, but they are predicted significantly higher in energy.

\begin{table}
\begin{tabular}{|l|l|l|}
\hline
$J^{\pi}$ & $E$ (EXP) & $E$(SM)  \\ \hline
$3/2^{-}_{1}$ & 0 & 0  \\ \hline
$1/2^{-}_{1}$ & 2.12469 & 1.5629  \\ \hline
$5/2^{-}_{1}$ & 4.44489 & 4.5909  \\ \hline
$3/2^{-}_{2}$ & 5.02031 & 4.5201  \\ \hline
$7/2^{-}_{1}$ & 6.7429 & 7.0646  \\ \hline
$1/2^{+}_{1}$ & 6.7918 & 7.1743  \\ \hline
$5/2^{+}_{1}$ & 7.28551 & 6.9531  \\ \hline
$3/2^{+}_{1}$ & 7.97784 & 8.0855  \\ \hline
$3/2^{-}_{3}$ & 8.5603 & 8.2309  \\ \hline
$5/2^{-}_{2}$ & 8.9202 & 8.5669  \\ \hline
$7/2^{+}_{1}$ & 9.185 & 8.8831  \\ \hline
$5/2^{+}_{2}$ & 9.2744 & 9.3921  \\ \hline
$1/2^{+}_{2}$ & 9.82 & 9.936  \\ \hline
$3/2^{+}_{2}$ & 9.876 & 10.6324  \\ \hline
$3/2^{-}_{4}$ & 10.26 & 10.3084  \\ \hline
$5/2^{-}_{3}$ & 10.33 & 10.2799 \\ \hline
$7/2^{+}_{2}$ & 10.597 & 10.3334  \\ \hline
$7/2^{-}_{2}$  &   & 10.7638  \\ \hline
$1/2^{-}_{2}$ &   & 10.8098  \\ \hline
$5/2^{-}_{4}$ & 10.96 & 12.2265  \\ \hline
$9/2^{+}_{1}$ & 11.265 & 10.6816 \\ \hline
$(3/2^{+}_{3})$ & 11.450 & 11.5736  \\ \hline
$9/2^{+}_{2}$  &   & 11.2997  \\ \hline
$5/2^{+}_{3}$ & 11.6 & 10.4898  \\ \hline
 $3/2^{-}_{5}$  &   & 11.3971  \\ \hline
$1/2^{-}_{3}$  &   & 11.6412  \\ \hline
$5/2^{-}_{5}$ & 11.886 & 12.8673  \\ \hline
$7/2^{+}_{3}$ & 12 & 12.4361  \\ \hline
$5/2^{+}_{4}$  &   & 12.3853  \\ \hline
$3/2^{+}_{4}$  &   & 12.4959  \\ \hline
$7/2^{+}_{4}$  &   & 12.9401  \\ \hline
$5/2^{+}_{5}$   &   & 13.0101  \\ \hline
$1/2^{+}_{3}$  &   & 13.2499  \\ \hline
$3/2^{+}_{5}$  &   & 13.502  \\ \hline
\end{tabular}
\caption{\label{tab:1} 
States in $^{11}$B as they appear in the shell model calculation are listed in the first column. The states are identified by spin, parity, and an additional sequential number.
The corresponding shell model excitation energy with {\it psdu} interaction Hamiltonian is shown in the third column.  
The corresponding available experimental excitation energy is presented in the second column. All energies are in units of MeV. 
A number of states predicted by the shell model have not yet been observed in experiment, those spaces are left blank. 
The spin and parity of 
state at 11.450 are not established experimentally but based on our systematics we identified it with the shell model state 
$3/2^{+}_{3}.$
}
\end{table}

In the following Tab. \ref{tab:2} the transition rates are summarized for the states of interest. 

\begin{table}
\begin{tabular}{|c|c|c|c|c|c|c|c|}
\hline
$J^{\pi}_i $& E(exp) & E(psdu)& E(fsu)& $B(GT)$ & $\log_{10}(ft)$ &  SF$_p$ &  SF$_\alpha$  \\ \hline
$1/2^{+}_{1}$ & 6.79180 &  7.1743 & 6.5998&0.009 &5.625 & 0.078 & 0.198 \\ \hline
$3/2^{+}_{1}$ & 7.97784 & 8.0855 &8.6337&0.000 & 8.286& 0.040 & 0.034 \\ \hline
$1/2^{+}_{2}$ & (9.820) & 9.936 &10.9678 &0.191 & 4.301& 0.175 & 0.170 \\ \hline
$3/2^{+}_{2}$ & 9.873 & 10.6324 & 10.6968 &0.001 &6.737& 0.057 & 0.002\\ \hline
$3/2^{+}_{3}$ & (11.450) & 11.5736 &10.8534 &0.615&3.792 & 0.008 & 0.038 \\ \hline
$3/2^{+}_{4}$ &  & 11.9769 & 11.9453 &1.731 &3.343& 0.012& 0.008\\ \hline
$1/2^{+}_{3}$ & & 12.7309 & 12.4382&0.659 & 3.762& 0.019 & 0.012\\ \hline
$3/2^{+}_{5}$ &  & 12.9830 &12.2265 &0.035 & 5.046& 0.003 & 0.023 \\ \hline
$3/2^{+}_{6}$ &  & 13.1816 & 12.9324 &0.222 & 4.235& 0.001 & 0.007 \\ \hline
$1/2^{+}_{4}$ &  & 13.1963 & 13.5568 &0.959 & 3.599& 0.057 & 0.004 \\ \hline
$3/2^{+}_{7}$ &  & 14.4333 & 13.3750 &0.037 & 5.011& 0.003 & 0.001 \\ \hline
$3/2^{+}_{8}$ &  & 14.9305& 14.43 & 0.008  & 5.674& 0.013 & 0.001 \\ \hline
$1/2^{+},\,3/2$ & 12.554 & 14.7199 &12.6631 &0.571 &3.195& 0.230 & 0.000 \\ \hline
\end{tabular}
\caption{\label{tab:2} For the lowest states potentially accessible by the Gamow-Teller transition we show 
experimental excitation energy, energy from the shell model with {\it psdu} Hamiltonian and with {\it fsu} Hamiltonian, 
in the second, third and fourth columns, respectively. All energies are in units of MeV. The following columns show:
$B(GT)$; the $\log_{10}(ft)$ evaluated with eq (\ref{eq:ft}) where for the isobaric analog state (last state in the table) the Fermi component with $B(F)=3$ is included; proton spectroscopic factors for the decay out of those states to the ground state of $^{10}$Be is shown in the seventh column; 
and the alpha spectroscopic factors are in the last column.
 }
\end{table}

\section{Sequential decay analysis}
In the classical sequential decay the first and the second decays are completely independent random processes. Thus, 
the lifetime is fully determined by the first decay; in our case this means that the beta decay of $^{11}$Be to a resonance in $^{11}$B should be relatively fast and consistent with the observed half-life $t_{{\rm Be}\rightarrow \beta p}=1\times 10^6$ s. It is also important that 
the second decay proceeds predominantly along the proton channel, so that the final state containing $^{10}$Be is reached. The latter
condition is particularly stringent because alpha decay channel is open.

In quantum machanics 
the  $1\rightarrow 2 \rightarrow 3$ process where state $1$ with energy $E_1$ decays to state $3$ with energy $E_3$ via an intermediate resonant state $2$  is described by a second order amplitude, Ref. \cite{volya:2014ad:proc}, 
\begin{equation}
A(\epsilon_1,\epsilon_2) = \frac{A_1(\epsilon_1) A_2(\epsilon_2)}{\epsilon_2-\left ((E_2-E_3)-\frac{i}{2} \Gamma_2(\epsilon_2)\right )}
\end{equation}
where $E_2$ and $\Gamma_2$ are energy and width of an intermediate state or resonance. 
The $\epsilon_1$ and $\epsilon_2$ are the decay energies of the first and the second steps, so total energy released $E=E_1-E_3=\epsilon_1+\epsilon_2.$ 
For our beta-delayed proton decay process only single amplitude (decay path) contributes leading to the partial decay width
\begin{equation}
\frac{d\Gamma(E)}{d\epsilon_1 d\epsilon_2} = 2\pi \delta (E-\epsilon_1-\epsilon_2) \left | A(\epsilon_1,\epsilon_2) \right |^2.
\label{eq:gsec}
\end{equation}
The amplitudes of the first (beta decay) and the second (proton decay) processes  
are energy-dependent and are related to the corresponding partial decay widths as 
\begin{equation}
\gamma_\beta(\epsilon)=2\pi | A_1(\epsilon)|^2,\quad \gamma_p(\epsilon)=2\pi | A_2(\epsilon)|^2.
\end{equation} 
The beta decay width $\gamma_\beta$ is given by (\ref{eq:gb}).
The width of the proton decay can be evaluated as $\gamma_p(\epsilon)={\rm SF}_p\, \gamma_{\rm ws}(\epsilon)$ where the proton spectroscopic
factor can be taken from Tab. \ref{tab:2}. The energy-dependent function $\gamma_{\rm ws}(\epsilon)$ is calculated using Woods-Saxon potential model with parameters from Ref. \cite{schwierz:2007:art}. 

%

The strength of the beta decay is typically assessed with $ft,$ using eq. (\ref{eq:ft}). 
Taking $E_1=0$ and $Q=-E_3.$ We can represent the integrated  sequential decay width (\ref{eq:gsec}) with a similar expression 
\begin{equation}
{\cal F} t=\frac{\cal T}{B(F)+\lambda^2 B (GT)},
\label{eq:ft_eff}
\end{equation}
that contains and effective energy dependent function 
\begin{equation}
{\cal F}(Q)=\int_0^Q  \frac{d\epsilon}{2\pi} \frac{f(\epsilon) \gamma_2(Q-\epsilon)}{(\epsilon + E_2)^2+ \Gamma^2_2(Q-\epsilon)/2}.
\label{eq:calf}
\end{equation}
If intermediate $E_2<0$ the sequential decay is open and integral in (\ref{eq:calf}) has a pole, so that if $\Gamma_2$ is very small 
${\cal F}(Q)=f(Q)$ and classical limit is recovered. 

For the analysis of eq. (\ref{eq:ft_eff}) we take a fixed time value 
$t\equiv t_{{\rm Be}\rightarrow \beta p}=1\times 10^6$ s which is prompted by the experiment. In the following Fig. \ref{fig:virft}  
we study 
$\log_{10}({\cal F} t)$ for $Q=0.2807$ MeV as a function of the position of intermediate proton resonance $E_2.$ 
From the values shown in Tab. \ref{tab:2}, that reflect the right side of eq. (\ref{eq:ft_eff}) we infer 
the limits on the $\log_{10}({\cal F} t)$ which constraints the position of the intermediate proton resonance $E_2.$
We take the width
\begin{equation}
\Gamma_2(\epsilon)=\gamma_p(\epsilon)+\gamma_\alpha
\end{equation}
where $\gamma_\alpha$ is assumed to be constant. 
The alpha threshold is far away and the $\gamma_\alpha$ has to be small to assure that proton is emitted most of the time and we reach 
in the final state in ${^{10}}$Be.  The alpha decay width can be evaluated using a potential model, similar to the proton decay. 
In this region of energy the potential-model limit for the width is between 1 and 2 MeV. Together with the spectroscopic factors we expect typical widths to range from between 1 keV and 200 keV.  The $3/2^+_{2}$ at 9.873 MeV of excitation is known to have alpha decay width of 109 keV, Ref. \cite{refsgaard:2019}.

\begin{figure}[h]
\begin{center}
\includegraphics[width=0.8\linewidth]{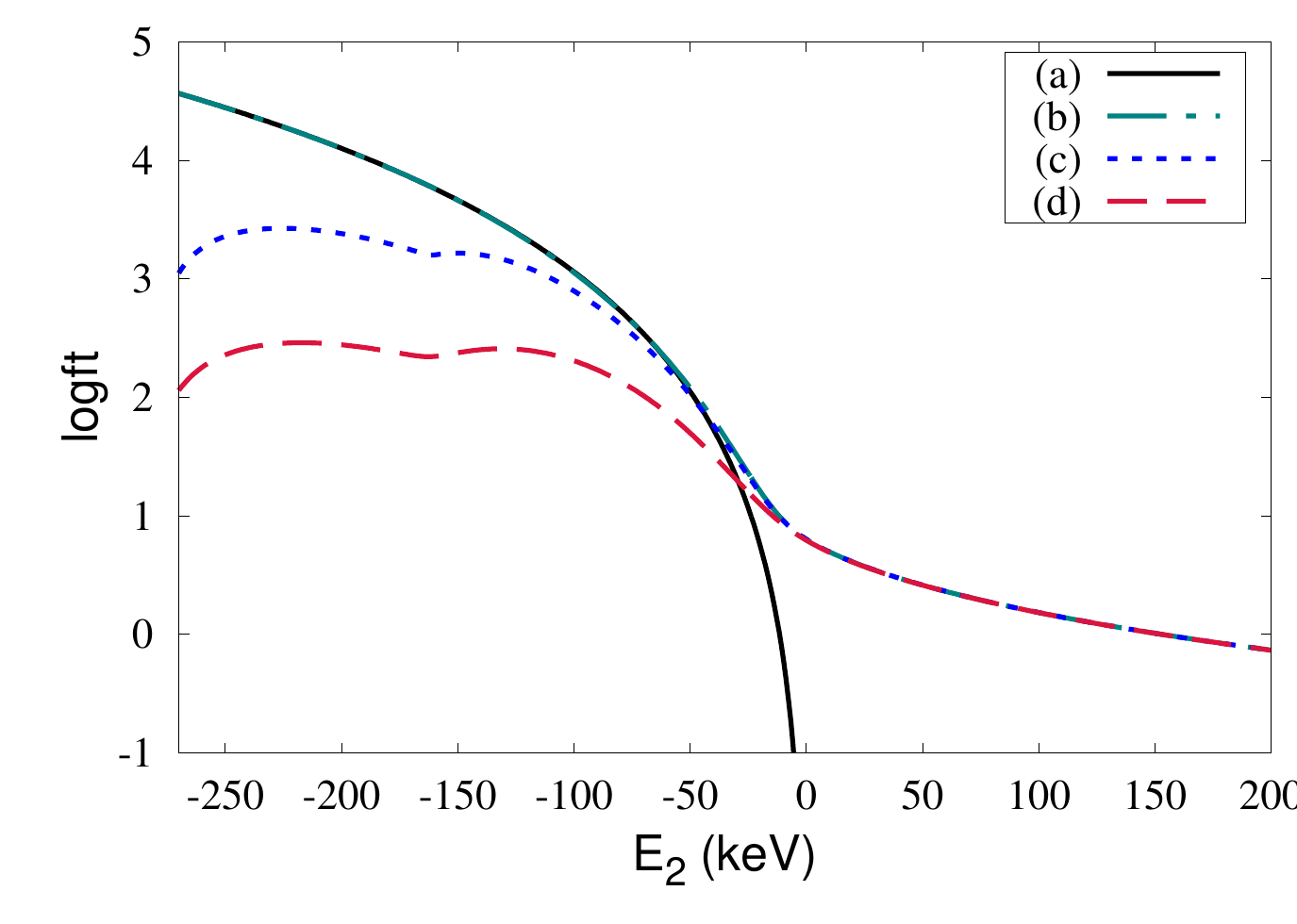}
\end{center}
\caption{\label{fig:virft} The evaluated for sequential decay $\log_{10}({\cal F} t)$ value for fixed $t\equiv t_{{\rm Be}\rightarrow \beta p}=1\times 10^6$ is shown as a function of $E_2$ which is the energy of the proton resonance in $^{11}$B. 
The energy $E_2$ is relative to the beta decay threshold. 
(a) Classical beta decay $\log_{10}(ft)$ for  $^{11}$Be decaying to $^{11}$B.
(b) Effective $\log_{10}({\cal F} t)$ for beta-delayed proton emission going through a resonant state in $^{11}$B, eq. (\ref{eq:ft_eff}),
with ${\rm SF}_p=0.23$ and  $\gamma_\alpha=0.$
(c) Same as (b) but with $\gamma_\alpha=1$ keV.
(d) Same as (b) and (c) but with $\gamma_\alpha=10$ keV.
}
\end{figure}

Let us start the discussion of Fig. \ref{fig:virft}  with a plot of $\log_{10}(f t)$ shown by a solid black line (a), this is a classical limit 
which is recovered when the width of an intermediate state is ignored $\Gamma_2=0.$ Considering the lowest
$\log_{10}(f t)$ listed in Tab. \ref{tab:2} we conclude that an intermediate state must be at least 150 keV below the threshold. 
This leaves at most 130 keV for proton decay, for $\epsilon_2<130$ keV $\gamma_{\rm ws}<2.3$ keV. Examining the spectroscopic factors in Tab. \ref{tab:2} the proton decay width at such energies should be less than about 0.2 keV.  In light of this, 
the $\gamma_p\approx 12$ keV and $\log_{10}(f t)\approx 4.8$ quoted in Ref. \cite{ayyad:2019} appear to be unrealistic. 
In addition, much larger alpha decay width further reduces the chances to see $^{10}$Be in the final state.

The most structurally favorable situation for the sequential decay process happens if an intermediate state is an isobaric analog state.
The alpha decay is isospin forbidden, the beta decay is enhanced by  the Fermi part, and the proton decay 
is favored by a large spectroscopic factor. This case is shown in Fig. \ref{fig:virft} with a dash-dot turquoise  line (b). 
In the region where $\log_{10}(f t)>3$ the result is indistinct from (a) meaning that full quantum mechanical consideration of the decay process  is unnecessary. Experimentally, the  isobaric analog state is believed to be about
1 MeV above the threshold, decay through such a virtual state is expected to be 4 to 5 orders of magnitude slower. 
This conclusion agrees with the phase space analysis in ref. \cite{ayyad:2019} suggesting $t_{{\rm Be}\rightarrow \beta p}= 2.2\times10^{10}$ s, and with the analysis in Ref.  \cite{baye:2011} taking into account additional spectroscopic suppressions from the shell model.

Finally, the blue dotted line (c) and long-dashed red line (d) both
show calculations that assume a large proton spectroscopic factor ${\rm SF}_p=0.23.$ and 
small alpha decay width $\gamma_\alpha=1$ and 10 keV, respectively. Since both curves never reach $\log_{10}(f t)>3.5$ it clear that even relatively weak alpha decay makes it completely impossible to explain the reported observations.  

\section{Conclusion}
The performed here analysis shows that it is very difficult to explain the reported in Ref. \cite{ayyad:2019} 
$t_{{\rm Be}\rightarrow \beta p}=1\times 10^6$ s half-life as being due to beta-delayed proton emission process. 
There are reports \cite{fynbo:2019,riisager:2020} suggesting problems with the experiment and highlighting some of the same issues.
In order to observe such rate, even for structurally favored superallowed states such as isobaric analog, the $Q$-value for the beta decay 
has to be over 100 keV, more likely over 150 keV. 
This leaves very little energy for the subsequent proton decay and the resulting proton 
decay width  is expected to be much less than a fraction of 1 keV. 
At the same time, all states in the region of interest are expected to have significant alpha decay width, likely about 100 keV, and 
certainly more than 1 keV.  The predominant alpha decay is expected to further reduce chances of seeing a beta-delayed proton decay 
by at least one or two orders of magnitude. 
The competition with alpha branch narrows the list of possible intermediate states in $^{11}$B to those with spin-parity $1/2^+.$
The shell model maybe rather uncertain in the prediction of energies but it describes well known states above and below the region of interest; and both versions considered predict nearest $1/2^+$ state about an MeV higher. 

We conclude that the reported half-life for the beta-delayed proton decay $t_{{\rm Be}\rightarrow \beta p}=1\times 10^6$ s is virtually impossible to explain with some of the best and most reliable theoretical models.  
Using Fig.~\ref{fig:virft} we estimate that even if there is an intermediate state that facilitates the 
beta-delayed proton decay it is unrealistic to expect the half-life shorter than $10^8$ s. However, it is most likely that such process 
proceeds virtually giving half-life of the order of $10^{10}$ s. Our evaluation of the principal contribution from the isobaric analog state gives $t_{{\rm Be}\rightarrow \beta p}=2.6\times 10^{10}$s, which is consistent with other theoretical estimates.

\acknowledgments
This material is based upon work supported by the US Department of Energy Office of Science, Office of Nuclear Physics under Award No. DE-SC0009883. Support from the Cyclotron Institute at Texas A\&M University is acknowledged. Discussions 
with V. Zelevinsky, G. Rogachev, V. Goldberg, J. Schiffer,  and M. Ploszajczak have been very helpful. 


%
%

\bibliographystyle{eplbib}

\end{document}